\documentclass[]{elsart}
\usepackage{amsmath,amsfonts}
 
\usepackage[dvips]{graphicx}
\usepackage{rotating}
 
\journal{}
 
\begin{document}
 
\begin{frontmatter}
 
\title{Can brane dark energy model be probed observationally by
distant supernovae?}
 
\author[WG,MS]{Marek Szyd{\l}owski}
and
\author[WG]{W{\l}odzimierz God{\l}owski}
 
\address[WG]{Astronomical Observatory, Jagiellonian University, Orla 171
30-224 Krak{\'o}w, Poland}
\address[MS]{Mark Kac Complex System Research Centre, Jagiellonian University,
Reymonta 4, 30-059 Krak{\'o}w, Poland}
 

\begin{abstract}
The recent astronomical measurements of distant supernovae as well as other
observations indicate that our universe is presently accelerating. There are
different proposals for the explanation of this acceleration, such as the
cosmological constant $\Lambda$, decaying vacuum energy, an evolving scalar
field (quintessence), phantom energy, etc. Most of these proposals require the
existence of exotic matter with negative pressure violating the strong energy
condition. On the other hand, there have appeared many models which offer
dramatically different mechanisms for the current acceleration, in which dark
energy emerges from the gravity sector rather than from the matter sector. In
this paper, we compare the concordance $\Lambda$CDM model with the
Sahni-Shtanov brane-world models of dark energy by using the Akaike and
Bayesian information criteria. We show that new parameters in the brane model
are not statistically significant in terms of the information criteria,
although the best fit method gives an improved fit to the SNIa data, because
of the additional parameters. This is because the information criteria of
model selection compensate for this advantage by penalizing models having more
free parameters. We conclude that only new future observational data are
accurate enough to give an advantage to dark-energy models of the brane origin,
i.e., a very high-significance detection is required to justify the presence
of new parameters. In our statistical analysis both Riess et al.'s and Astier
et al.'s SNIa samples are used. For stringent constraining parameters of the
models the baryon oscillation peak (BOP) test is used.
\end{abstract}
 
\end{frontmatter}

\section{Introduction}
 
The recent supernovae SNIa measurements \cite{Riess:1998cb,Perlmutter:1998np}
as well as other observations indicate that the expansion of our present
universe is accelerating. While the cosmological constant offers the
possibility of effective explanation of the acceleration, the existence of
fine tuning difficulties motivate theorist to search for alternative forms
of dark energy. All these proposals can be divided into two groups following
the criterion whether dark energy emerges from gravity or matter sectors of
the theory. The first group is characterized by postulating the existence of
unusual properties of matter content with negative pressure violating the
strong energy condition. This category of models includes (besides the
$\Lambda$CDM model or varying $\Lambda$-term model) the quintessence models
(referred to by some as models of an evolving scalar field), phantom models,
etc. As a representative model of this class, we consider the $\Lambda$CDM
model which we confront with a subclass of models of the second category---the
brane models of dark energy. In these models, dark energy emerges from a
different evolutional scenario at the late time of evolution. In this approach,
instead of a new hypothetical energy component of an unknown form, dark energy
arises from the modified gravity sector of the theory. The basic idea in these
cosmologies is that our observable Universe is a four-dimensional brane
embedded in a five-dimensional bulk space. As the representatives of this
cosmologies, we consider two classes of models which appeared in recent
achievement, namely (i)~the Deffayet-Dvali-Gabadadze (DDG) model
\cite{Dvali:2000hr,Dvali:2000xg,Deffayet:2000uy} and (ii)~the Sahni-Shtanov
(SSh) model \cite{Shtanov:2003,Sahni:2003}. The main difference between these
two models is that in the second one includes both brane and bulk cosmological
constants and, similarly to the DDG model, also includes the scalar curvature
term in the action for the brane. The Randall-Sundrum (RS) model
\cite{Randall:1999ee,Randall:1999vf} can be recovered as a special limit of the
SSh model. Thus, the SSh model generalizes both the RS model and the DDG model.
Compared to general relativity, both the DDG and SSh models introduce some
extra parameters, and then it is crucial to perform an objective comparison of
these models. In the generic case, introducing extra parameters result
naturally in an improved fit to the data, but a crucial question is whether
these new parameters are actually relevant for explaining SNIa data set
\cite{Liddle:2004nh}.
 
One of the most popular procedures adopted to compare models with a different
parameters is to use the best-fit method based on the maximum of the
likelihood function. However, it is well known that this method favours the
model with the largest number of parameters. The likelihood ratio test
\cite{Kendal} based on the simplest procedure of calculating the quantity
$2\ln{\frac{\mathcal{L_{\text{simple}}}}{\mathcal{L_{\text{complex}}}}}$
(where $\mathcal{L}$ is the maximum likelihood) can also be useful in this
context. While this ratio can be used to control the significance of any
increase in the likelihood against an additional parameter introduced to
the model, the assumptions of this criterion are often violated in the
astrophysical applications \cite{Protasov:2002}.
 
The key aim of this letter is to make an objective comparison of two different
groups of dark energy models which may feature different numbers of model
parameters. We use the Akaike information criteria (AIC) \cite{Akaike:1974} and
the Bayesian information criteria (BIC) \cite{Schwarz:1978} to select model
parameters providing the preferred fit to data. These information criteria
enable us to select the combination of cosmological parameters giving
the best fit to the present SNIa data. Of course, some future observational
data (from SNAP, for example) may give arguments in favour of additional
parameters of brane-world dark energy, but here we claim that such models
have no impact on the current Universe. Taking into account the simplicity
argument, we argue that for the verification of the idea of brane-world dark
energy very high significance detection is required and, therefore, at present
these extra parameters have a marginal significance in the fits to the
present data.

\section{Cosmologies with brane dark energy origin}
 
It was pointed out by many authors \cite{Sahni:2002kh,Sahni05a,Brax2004} that
brane models offer a wider range of possibilities for solving the problem of
acceleration than standard $\Lambda$CDM model. Alam and Sahni \cite{Alam02}
claimed that the ``Brane (1)'' model, which has the effective equation of
state $w \equiv p/\rho <-1$, provides better agreement with the SNIa data than
the $\Lambda$CDM model for matter density parameter $\Omega_{\text{m},0}>0.3$
(and for $\Omega_{\text{m},0} \le 0.25$). Such a conclusion comes from a
simple comparison of the best-fit method based on the maximum likelihood
function which usually favour models with the largest number of parameters, in
our case, the SSh model over the $\Lambda$CDM model. In Table~\ref{tab:1}, two
different brane models and the reference $\Lambda$CDM model are represented
in terms of the Hubble parameter $H$ as a function of redshift $z$. For
simplicity we considered the flat case $\Omega_{k,0}=0$ which is strongly
preferred by the WMAP data \cite{Bennett:2003bz}. Also in our previous paper
\cite{Godlowski:2005tw,Szydlowski:2005} we find that, when we analyze fit to
SNIa data, in general cases the number of essential parameters in the cosmological
models with dark energy is in principal, equal to two, namely, $H_0$,
$\Omega_{\text{m},0}$, i.e., $\Omega_{k,0}=0$ is not an essential parameter.
 
The presence of two bulk and brane cosmological constants distinguishes the SSh
model from the DDG model (see Table~\ref{tab:1}). In the terminology of Sahni
and Shtanov the SSh model is called the Brane (1) model according to different
ways of bounding anti de Sitter (or Schwarzschild) bulk space by the brane
\cite{Sahni05}). The decaying $C/a^4$ dark radiation term can be neglected in
the basic Friedmann equation \cite{Shtanov02}. The generalized Friedmann
equation assumes the following form
\begin{equation} \label{eq:1}
H^2+\frac{k}{a^2}=\frac{\Lambda_{\mathrm{b}}}{6}+\frac{C}{a^4}+
\frac{1}{l^2}\left(\sqrt{1+l^2\left(\frac{\rho+\sigma}{3m^2}-
\frac{\Lambda_{\mathrm{b}}}{6}-\frac{C}{a^4}\right)} \mp 1\right)^2
\end{equation}
where $l=\frac{2m^2}{M^2}$ is the length scale, $m^2$ is the coupling constant
in action $M$ and $m$ denote the five and four dimensional Planck masses
respectively, $\Lambda_{\text{b}}$ is the cosmological constant on the bulk,
$\sigma$ is the brane tension, and the plus/minus sign before the last term
corresponds to ``Brane (1)'' and ``Brane (2)'' solution, respectively.
 
In Table~\ref{tab:1}, equation (\ref{eq:1}) is expressed in terms of
dimensionless density parameters for the flat case
$\Omega_{\text{m},0} = \frac{\rho_0}{3m^2H_0^2}$,
$\Omega_{\sigma,0} = \frac{\sigma}{3m^2H_0^2}$,
$\Omega_{\text{l},0} = \frac{1}{l^2H_0^2}$,
$\Omega_{\Lambda_{\text{b}},0} = -\frac{\Lambda_b}{6H_0^2}$.
 
For completeness we consider the flat brane cosmology which bases on
Deffayet's modification of the FRW equation \cite{Deffayet:2000uy,Lue04}
\begin{equation}
\label{eq:1a}
H^2 \pm \frac{H}{r_0}=\frac{\rho}{3}+\frac{\Lambda}{3}
\end{equation}
where $r_0=\mathcal{M}_{Pl}^2/2M^3$ is the scale on which it is possible to
``probe'' the extra dimension, the plus/minus sign corresponds the two distinct
cosmological phases---the self accelerating and Minkowski cosmological ones.
This model, which fits well to SNIa data, was analysed by Lue and Starkman \cite{Lue04}.
This model belongs to a class of models can be derived from the SSh models
after setting the bulk cosmological constant to zero. Therefore these models are
classified as SSh($\Lambda_b=0$).
 
In Table~\ref{tab:1} we complete all these models together with the dependence
of Hubble's function $H=\frac{\dot{a}}{a}$ on redshift. We also denote the
number of a model's independent, free parameters by $d$. Note that we have
constraint on all rewritten $\Omega_{i,0}$ parameters from the condition
$H(z=0) = H_{0}$.
 
\begin{sidewaystable}
\begin{tabular}{c|p{2.5cm}|llc}
\hline
case & model   & $\qquad H(z)$ & model parameters & $d$ \\
\hline
1  & $\Lambda$CDM   & $H=H_0 \sqrt{\Omega_{\text{m},0}(1+z)^3+\Omega_{\Lambda,0}}$ & $H_0,\Omega_{\text{m},0},\Omega_{\Lambda,0}$ & 2 \\
2  & DDG & $H=H_0 \sqrt{\left(\sqrt{\Omega_{\text{m},0}(1+z)^3+\Omega_{\text{rc},0}}+\sqrt{\Omega_{\text{rc},0}} \right)^2}$ & $H_0,\Omega_{\text{m},0},\Omega_{\text{rc},0}$ & 2 \\
3a & SSh Brane 1& $H=H_0 \sqrt{\Omega_{\text{m},0}(1+z)^3+\Omega_{\sigma,0}+2\Omega_{\text{l},0} - 2\sqrt{\Omega_{\text{l},0}}\sqrt{\Omega_{\text{m},0}(1+z)^3+\Omega_{\sigma,0}+\Omega_{\text{l},0}+\Omega_{\Lambda_{\text{b}},0}}}$
   & $H_0,\Omega_{\text{m},0},\Omega_{\sigma,0},\Omega_{\text{l},0},\Omega_{\Lambda_{\text{b}},0}$& 4\\
3b & SSh Brane 2& $H=H_0 \sqrt{\Omega_{\text{m},0}(1+z)^3+\Omega_{\sigma,0}+2\Omega_{\text{l},0} + 2\sqrt{\Omega_{\text{l},0}}\sqrt{\Omega_{\text{m},0}(1+z)^3+\Omega_{\sigma,0}+\Omega_{\text{l},0}+\Omega_{\Lambda_{\text{b}},0}}}$
   & $H_0,\Omega_{\text{m},0},\Omega_{\sigma,0},\Omega_{\text{l},0},\Omega_{\Lambda_{\text{b}},0}$& 4\\
4a & SSh1($\Lambda_b=0$)& $H=H_0 \sqrt{\Omega_{\text{m},0}(1+z)^3+\Omega_{\sigma,0}+2\Omega_{\text{l},0} - 2\sqrt{\Omega_{\text{l},0}}\sqrt{\Omega_{\text{m},0}(1+z)^3+\Omega_{\sigma,0}+\Omega_{\text{l},0}}}$
   & $H_0,\Omega_{\text{m},0},\Omega_{\sigma,0},\Omega_{\text{l},0}$& 3\\
4b & SSh2($\Lambda_b=0$)& $H=H_0 \sqrt{\Omega_{\text{m},0}(1+z)^3+\Omega_{\sigma,0}+2\Omega_{\text{l},0} + 2\sqrt{\Omega_{\text{l},0}}\sqrt{\Omega_{\text{m},0}(1+z)^3+\Omega_{\sigma,0}+\Omega_{\text{l},0}}}$
   & $H_0,\Omega_{\text{m},0},\Omega_{\sigma,0},\Omega_{\text{l},0}$& 3\\
\hline
\end{tabular}
\caption{The Hubble function and parameters for different models explaining
acceleration in terms of dark energy.} \label{tab:1}
\end{sidewaystable}

\section{The Akaike and Bayesian information criteria}
 
The information criteria (in a similar way as the adjusted coefficient of
determination in standard statistics) put a threshold which must be
exceeded in order to assert an additional parameter to be important
in explanation of the phenomenon. The discussion how high this threshold
should be caused appearing many different criteria. The Akaike and Bayesian
information criteria (AIC and BIC) (for review see \cite{BA2004}) are most
popular and used in everyday statistical practices.
 
In the case of the model in question we find that the AIC and BIC
information criteria of model selection do not provide
sufficient arguments for incorporation of new parameters from brane
cosmology when SNIa data are used. It is in contrast to the conclusion
which Alam and Sahni \cite{Alam02,Alam05}) obtained without using the
information criteria.
 
The usefulness of using information criteria of model selection was recently
demonstrated by Liddle \cite{Liddle:2004nh} and Parkinson et al.
\cite{Parkinson:2005}
 
The AIC is defined in the following way \cite{Akaike:1974}
\begin{equation} \label{eq:11}
\text{AIC} = - 2\ln{\mathcal{L}} + 2d
\end{equation}
where $\mathcal{L}$ is the maximum likelihood and $d$ is a number of the model
parameters. The best model with a parameter set providing the preferred fit to
the data is that minimizes the AIC. It is interesting that the AIC also
arises from an approximate minimization of the Kulbak-Leibner information
entropy \cite{Sakamoto86}.
 
The BIC introduced by Schwarz \cite{Schwarz:1978} is defined as
\begin{equation} \label{eq:12}
\text{BIC} = - 2\ln{\mathcal{L}} + d\ln{N}
\end{equation}
where $N$ is the number of data points used in the fit. The AIC tends to
favour models with large number of parameters when compared to the BIC, so the
latter provides a more useful approximation to the full statistical analysis
in the case of no priors on the set of model parameters \cite{Parkinson:2005}.
It makes this criterion especially suitable in context of cosmological
applications.
 
It is pointed out that while the AIC is useful in obtaining upper limit to the
number of parameters which should be incorporated to the model, the BIC is
more conclusive. Of course only the relative value between BIC of different
models has statistical significance. The difference of $2$ is treated as
a positive evidence (and $6$ as a strong evidence) against the model with a
larger value of the BIC \cite{Jeffreys:1961,Mukherjee:1998wp}.
If we do not find any positive evidence from information criteria the
models are treated as a identical and eventually additional parameters are
treated as not significant. The using of
the BIC seems to be especially suitable whenever the complexity of reference
does not increase with the size of data set which is important in the context
of the future SNAP observations. In a footnote Liddle \cite{Liddle:2004nh}
noted that in cosmology, a new parameter is usually a quantity set to zero in
a simpler base model and if the likelihood function is a continuous function
of its parameters it will increase as the parameter varies in either the
positive or negative direction.

\section{Distant supernovae as cosmological probes dark energy origin}
 
In this section it will be demonstrated that using the Akaike and Bayesian
information criteria one can answer which cosmological model is favoured?
We consider only two types of models 1) a model with dark energy violating
strong energy condition or 2) a model in which dark energy has brane origin.
We use the ``Gold'' SNIa data set selected by Riess et al. \cite{Riess:2004nr}.
This sample contains 157 SNIa with redshift up to $z=1.75$,
Recently Astier et al. \cite{Astier:2005} have compiled a new sample of 115
supernovae based on 71 high redshift type Ia supernovae discovered during
the first year of the 5-year Supernovae Legacy Survey. Thanks to the multi-band,
rolling search technique and careful calibration, this data set is arguably
the best high-z SNIa compiled data \cite{Astier:2005,Fairbairn:2005},
unfortunately only with supernovae up to $z=1$. It is the main reason that we
decide to use both SNIa samples in our analysis.
 
For the distant SNIa one can directly observe their apparent magnitude $m$
and redshift $z$. Because the absolute magnitude $\mathcal{M}$ of the
supernovae is related to its absolute luminosity $L$, then the relation between
the luminosity distance $d_L$ and the observed magnitude $m$ and the absolute
magnitude $M$ has the following form
\begin{equation} \label{eq:4}
m - M = 5\log_{10}d_{L} + 25.
\end{equation}
Instead of $d_L$, the dimensionless parameter $D_L$
\begin{equation} \label{eq:5}
D_{L}=H_{0}d_{L}
\end{equation}
is usually used and then eq.~(\ref{eq:4}) changes to
\begin{equation} \label{eq:6}
\mu \equiv m - M = 5\log_{10}D_{L} + \mathcal{M}
\end{equation}
where
\begin{equation} \label{eq:7}
\mathcal{M} = - 5\log_{10}H_{0} + 25.
\end{equation}
We know the absolute magnitude of SNIa from the light curve. The luminosity
distance of a supernova can be obtain as the function of redshift
\begin{equation} \label{eq:8}
d_L(z) =  (1+z) \frac{c}{H_0} \frac{1}{\sqrt{|\Omega_{k,0}|}} \mathcal{F}
\left( H_0 \sqrt{|\Omega_{k,0}|} \int_0^z \frac{d z'}{H(z')} \right)
\end{equation}
where $\Omega_{k,0} = - \frac{k}{H_0^2}$ and
\begin{align}
\label{eq:9}
\mathcal{F} (x) &=  \sinh (x) \qquad &\text{for} &\qquad k<0 \nonumber \\
\mathcal{F} (x) &=         x  \qquad &\text{for} &\qquad
  k=0    \\
\mathcal{F} (x) &=  \sin (x)  \qquad &\text{for} &\qquad k>0. \nonumber
\end{align}
 
Finally, it is possible to probe dark energy which constitutes the main
contribution to the matter content. It is assumed that supernovae measurements
come with the uncorrelated Gaussian errors and in this case the likelihood
function $\mathcal{L}$ can be determined from the chi-square statistic
$\mathcal{L}\propto \exp(-\chi^{2}/2)$ where
\begin{equation} \label{eq:10}
\chi^{2}=\sum_{i}\frac{(\mu_{i}^{\mathrm{theor}}-\mu_{i}^{\mathrm{obs}})^{2}}
{\sigma_{i}^{2}},
\end{equation}
($\sigma_i$ denotes the full statistical error of magnitude determination
including error in $z$ measurement)
while the probability density function of cosmological parameters
\cite{Riess:1998cb} is derived from Bayes' theorem. Therefore, we can
perform the estimation of the model's parameters using the minimization
procedure, based on the likelihood function.
 
The results of statistical calculation of considered dark energy models are
presented in Table~\ref{tab:2} and Table~\ref{tab:3}. In Table~\ref{tab:2} we
show results for models considered (flat cases) without any assumed extra
priors. In Table~\ref{tab:3} we presented analogous results for the case
of the assumed prior $\Omega_{\text{m},0}=0.3$ \cite{Peebles:2002gy}.
Since the value of matter density is not yet known precisely
we also present results for $\Omega_{\text{m},0}=0.2$ and
$\Omega_{\text{m},0}=0.4$.

Using the best fit method based on the maximum of the likelihood function
(minimum $\chi^2$) we conclude that SSh models are a better fit than the
$\Lambda$CDM model if we consider models without any assumed extra priors.
 
\begin{figure}
\includegraphics[width=0.5\textwidth]{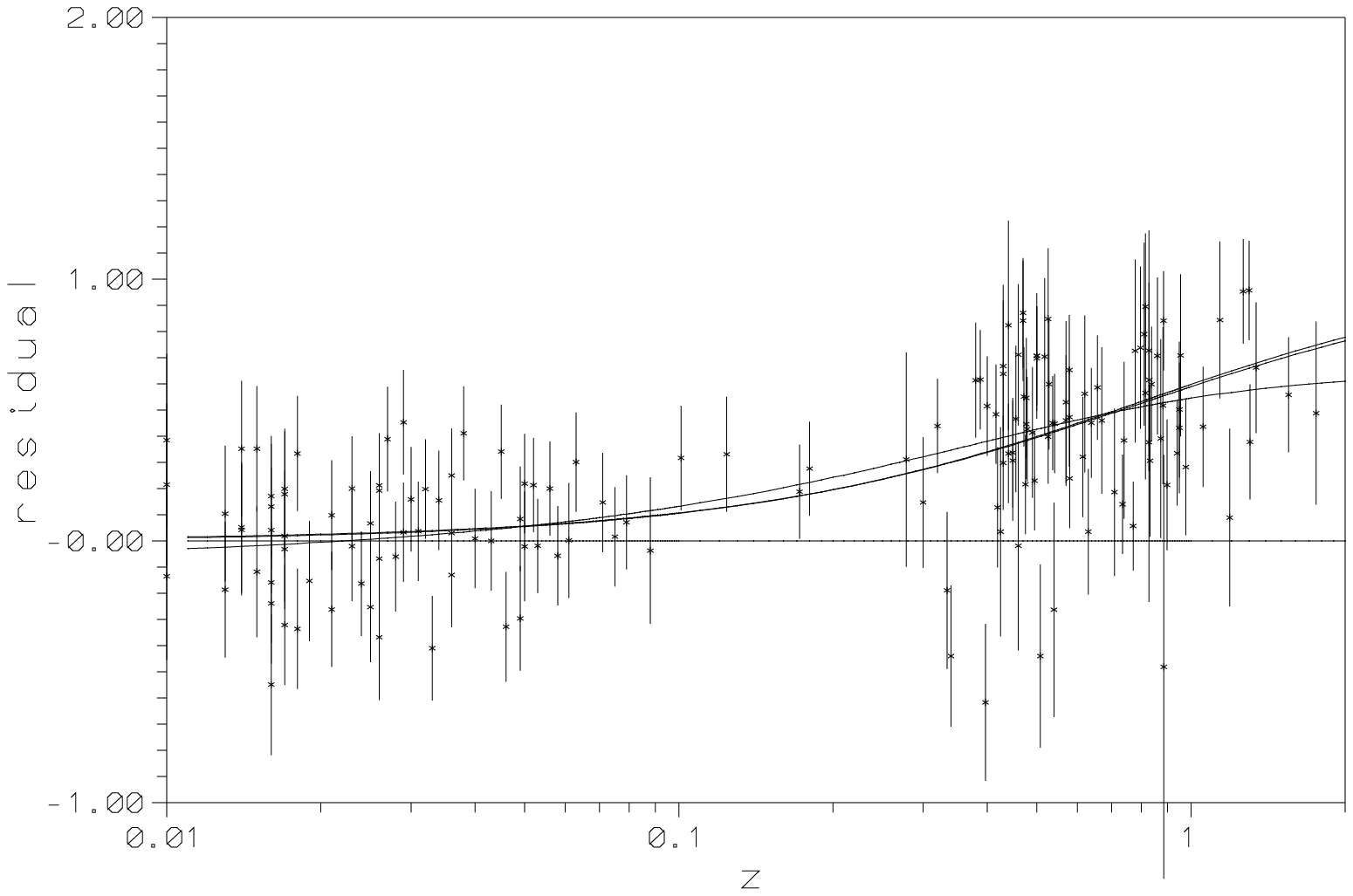}
\includegraphics[width=0.5\textwidth]{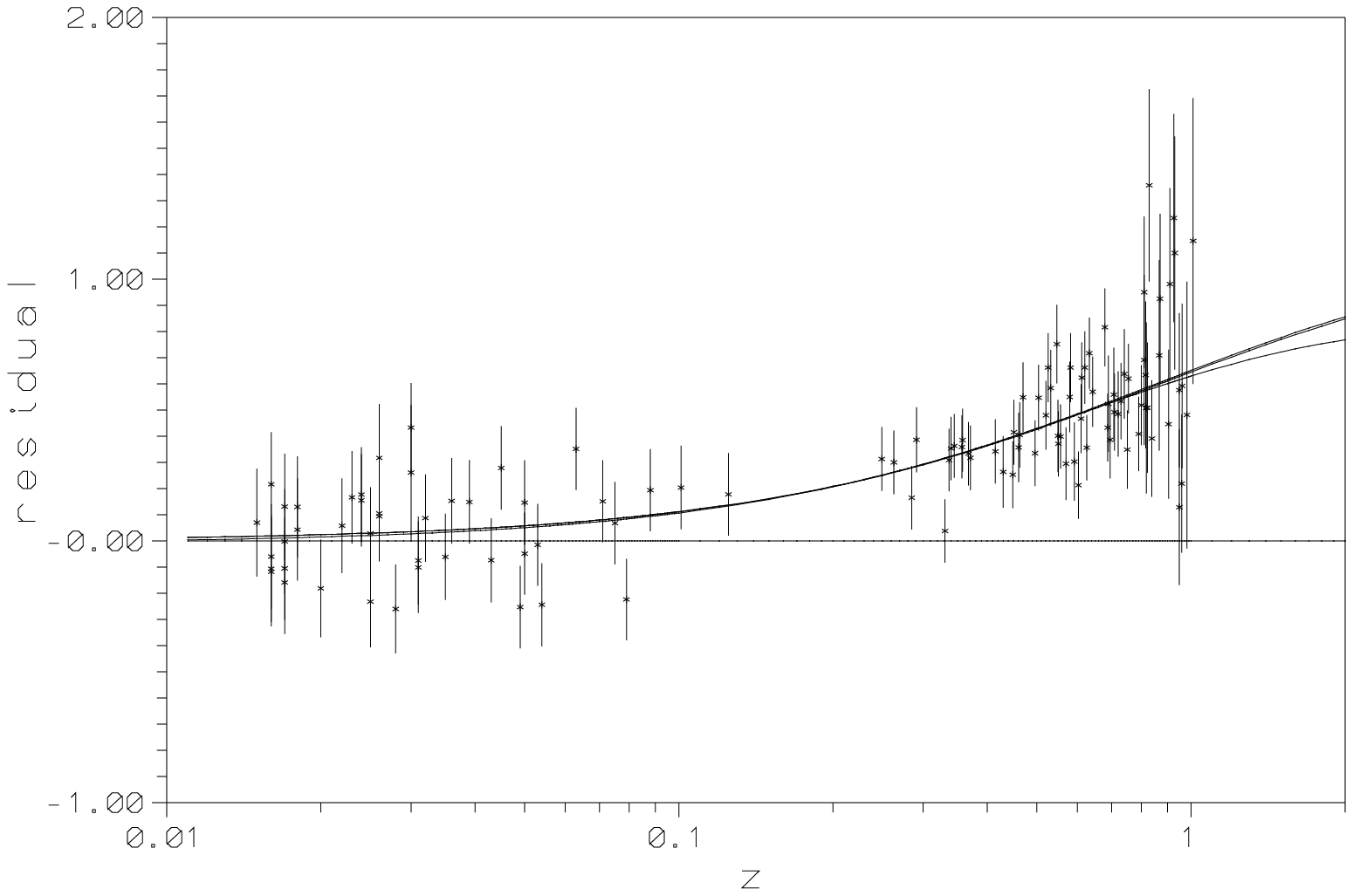}
\caption{Residual plots of the redshift-magnitude relations between the
Einstein-de Sitter model (represented by the zero line) the best-fitted SSh
``Brane (1)'' model (middle curve) and the flat $\Lambda$CDM model---the
upper curve. Best fitted SSh ``Brane 2'' model is inseparable from
the $\Lambda$CDM model. Left panel was obtain with the Gold sample while
the right panel was obtained with Astier et al.'s sample.}
\label{fig:1}
\end{figure}

The above mentioned results are illustrated in Fig.~\ref{fig:1}. We present
residual plots of the redshift-magnitude relations between the Einstein-de
Sitter model (represented by zero line) the best-fitted SSh ``Brane(1)''
model (middle curve) and the flat $\Lambda$CDM model---the upper curve.
Please note that best fitted ``Brane (2)'' model is inseparable from the
$\Lambda$CDM model. These two models seems to be inseparable in
the low and middle redshifts (see also \cite{Sahni05}). One can observe that
the systematic deviation between the $\Lambda$CDM model and the SSh ``Brane(1)''
model gets larger at higher redshifts ($z>0.9$). The SSh ``Brane(1)'' model
predicts that high redshift supernovae should be brighter than predicted with
the $\Lambda$CDM model.
 
\begin{center}
\begin{figure}
\includegraphics[width=0.5\textwidth]{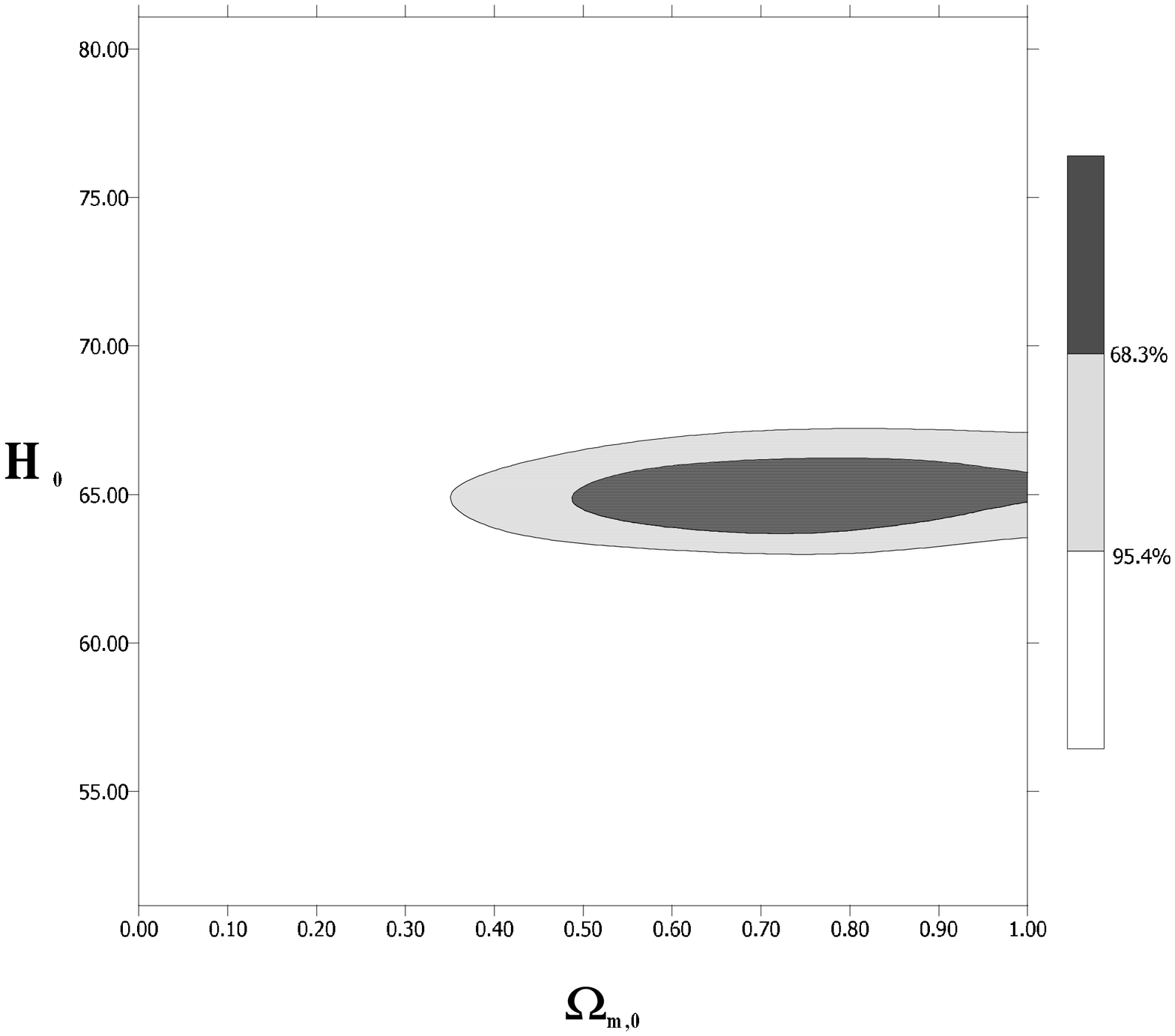}
\includegraphics[width=0.5\textwidth]{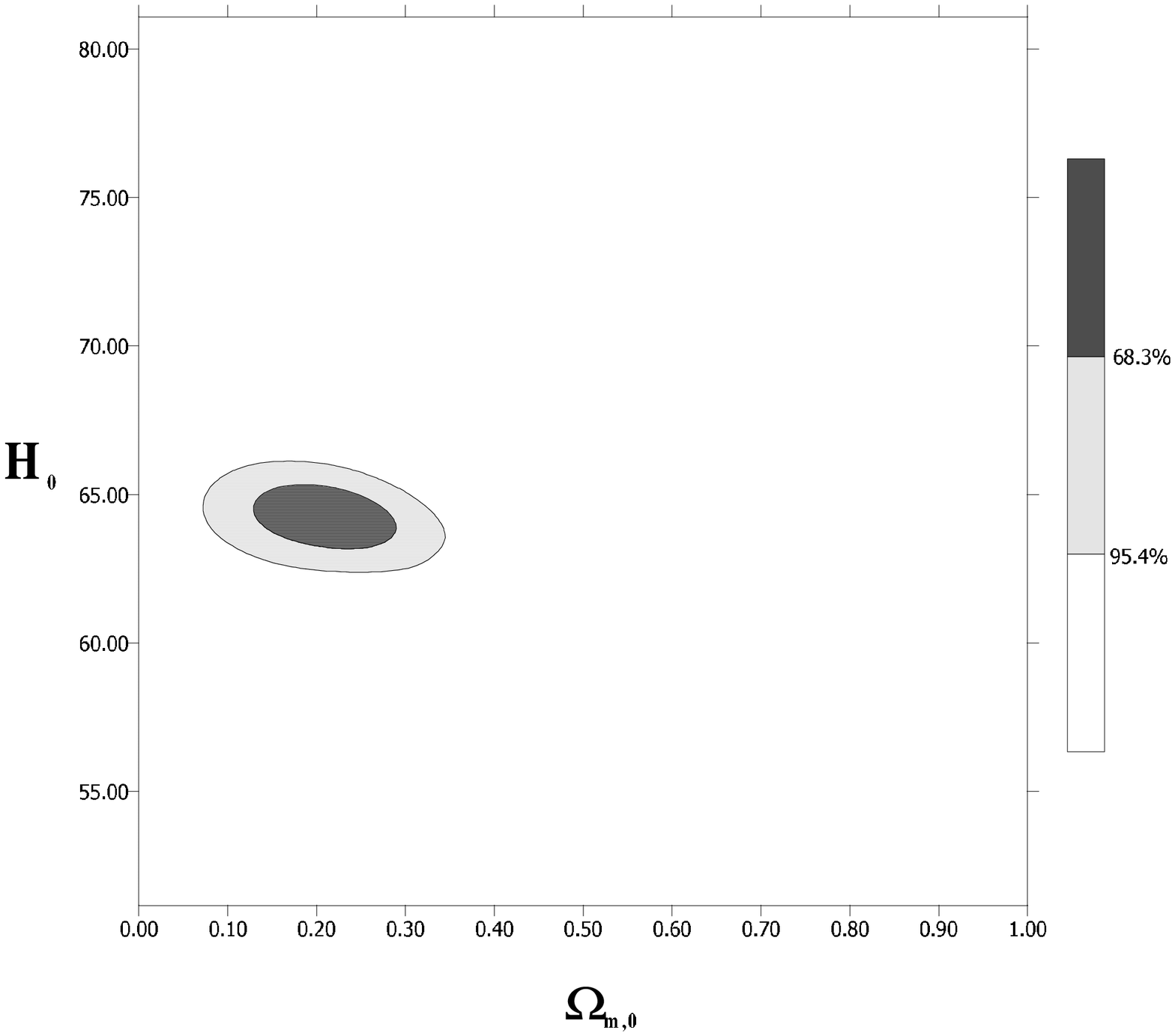}
\caption{Confidence levels intervals for the SSh ``Brane (1)'' (left panel)
and ``Brane (2)'' (right panel) models on the $(\Omega_{\text{m},0}, H_0)$
plane. They are obtained with the Gold sample by marginalizing the probability
density functions over remaining parameters (assuming uniform priors).}
\label{fig:2}
\end{figure}
\end{center}

We supplement our analysis of the SSh model with confidence levels intervals
in the $(\Omega_{\text{m},0}, H_0)$ plane by marginalizing the probability
density functions over remaining parameters assuming uniform priors.
(Fig.~\ref{fig:2}). We obtain that the SSh ``Brane (1)'' model prefers
universes with high density of $\Omega_{\text{m},0} \simeq 0.75$, while
the SSh ``Brane (2)'' model prefers universes with low density of
$\Omega_{\text{m},0} \simeq 0.20$. It is a situation contrary to the
$\Lambda$CDM where $\Omega_{\text{m},0} \simeq 0.3$ is preferred
\cite{Riess:1998cb,Perlmutter:1998np,Riess:2004nr}.
 
The results of the AIC and BIC in the context of considered models
(Table~\ref{tab:1}) are collected in Tables~\ref{tab:2} and \ref{tab:3}.
One can observe that the BIC as well as AIC values assumes lower values
for the $\Lambda$CDM model when we do not assumed any prior for
$\Omega_{m,0}$.
 
\begin{figure}
\includegraphics[width=0.45\textwidth]{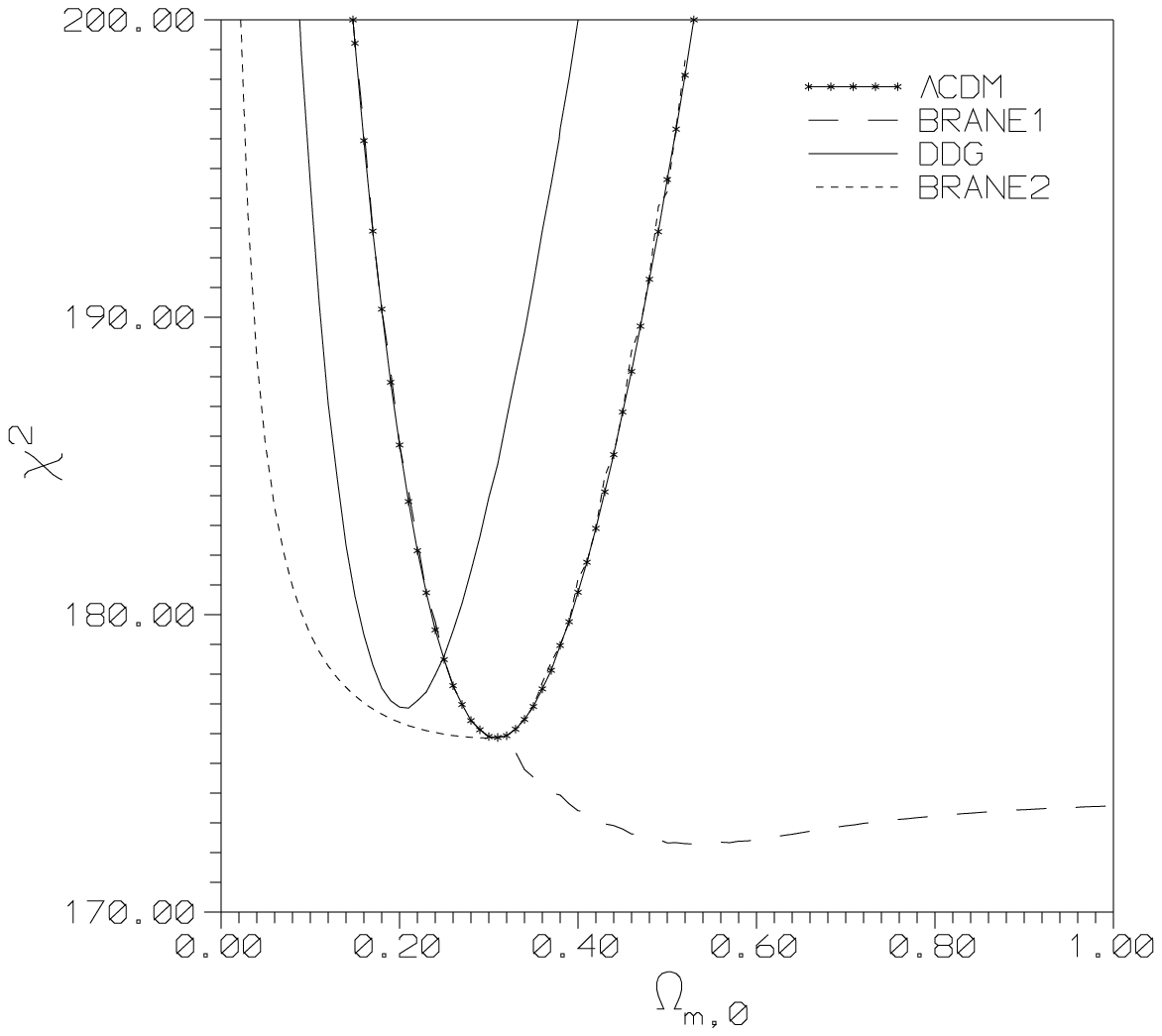}
\includegraphics[width=0.45\textwidth]{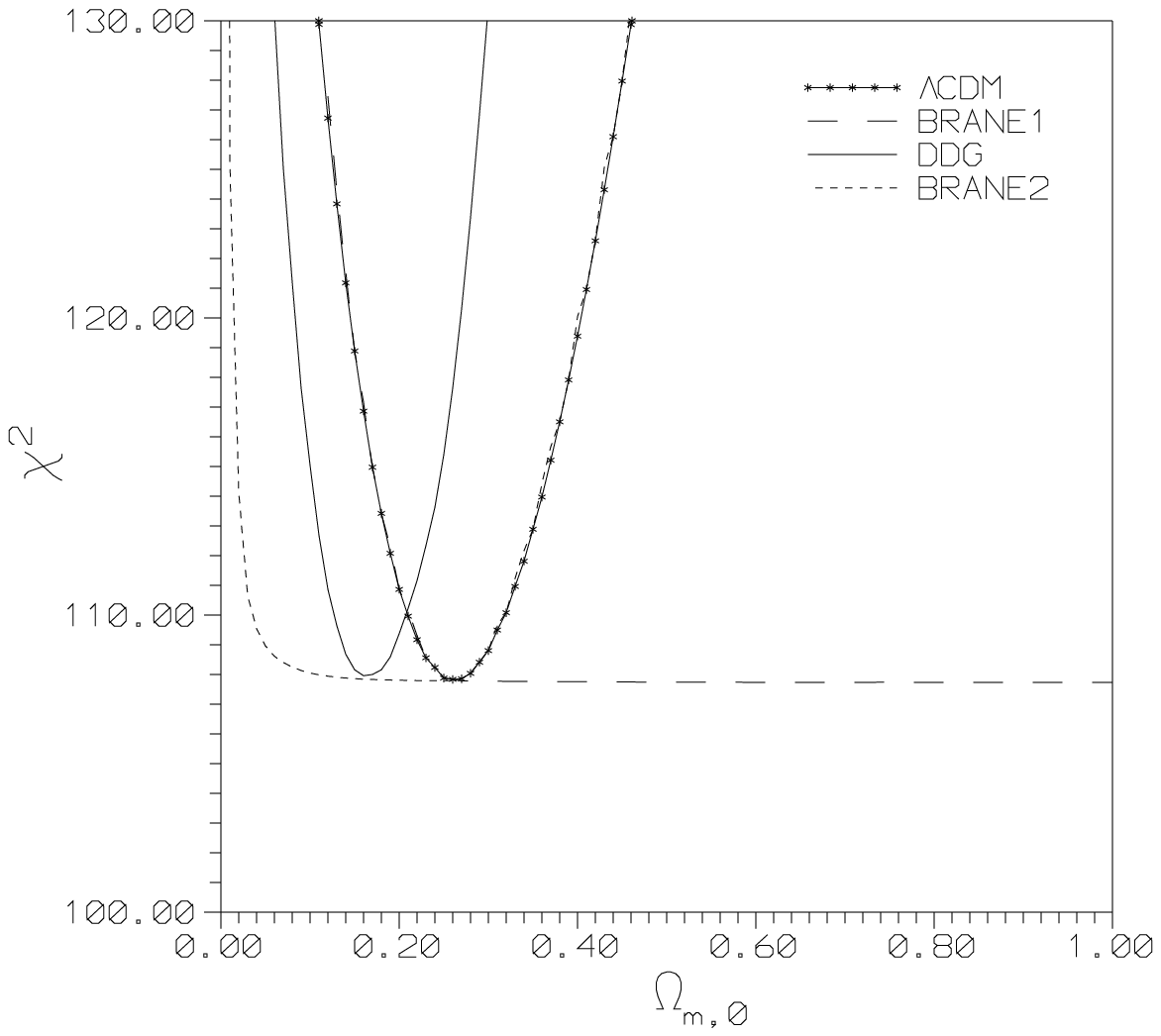}\\
\includegraphics[width=0.45\textwidth]{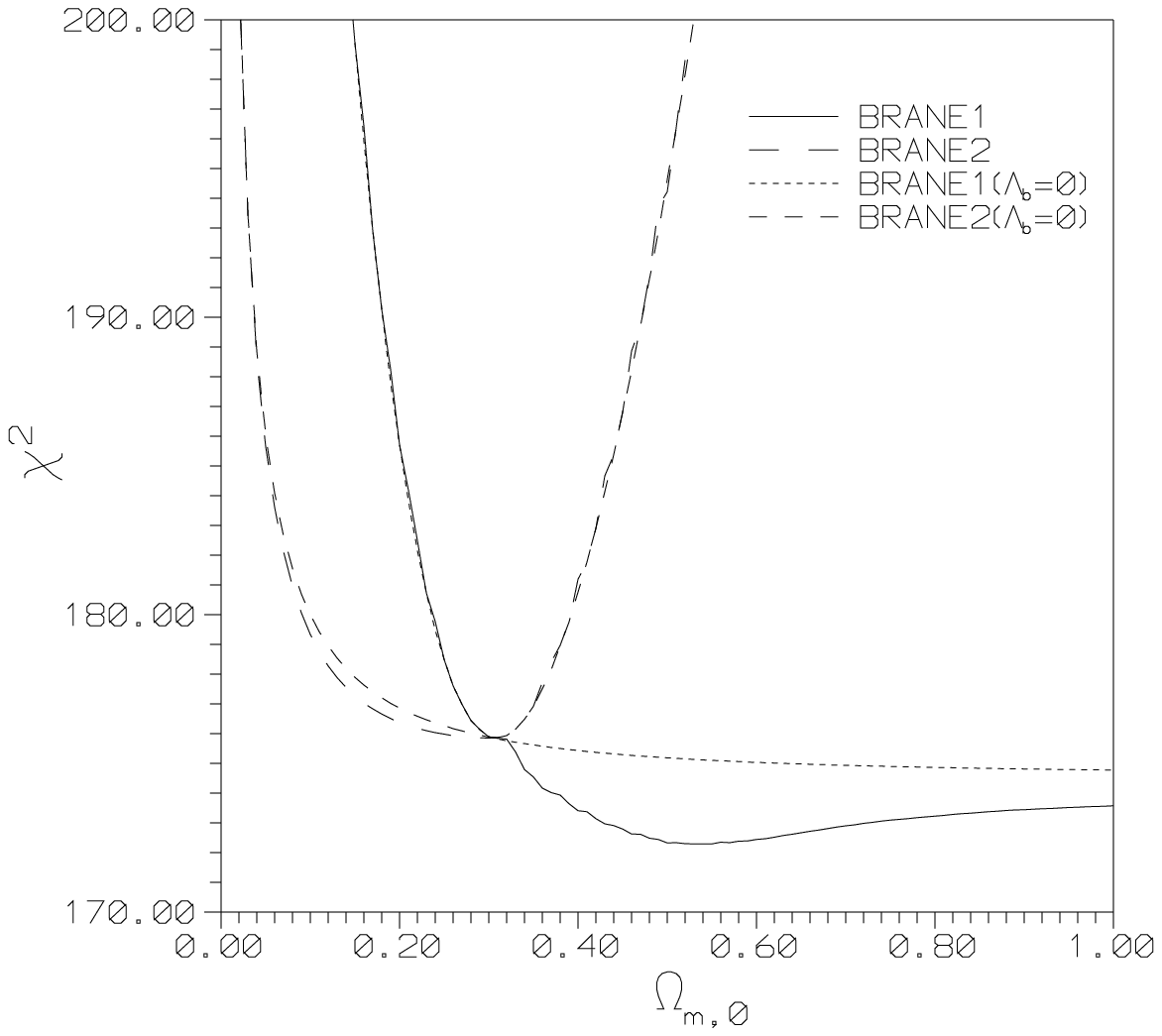}
\includegraphics[width=0.45\textwidth]{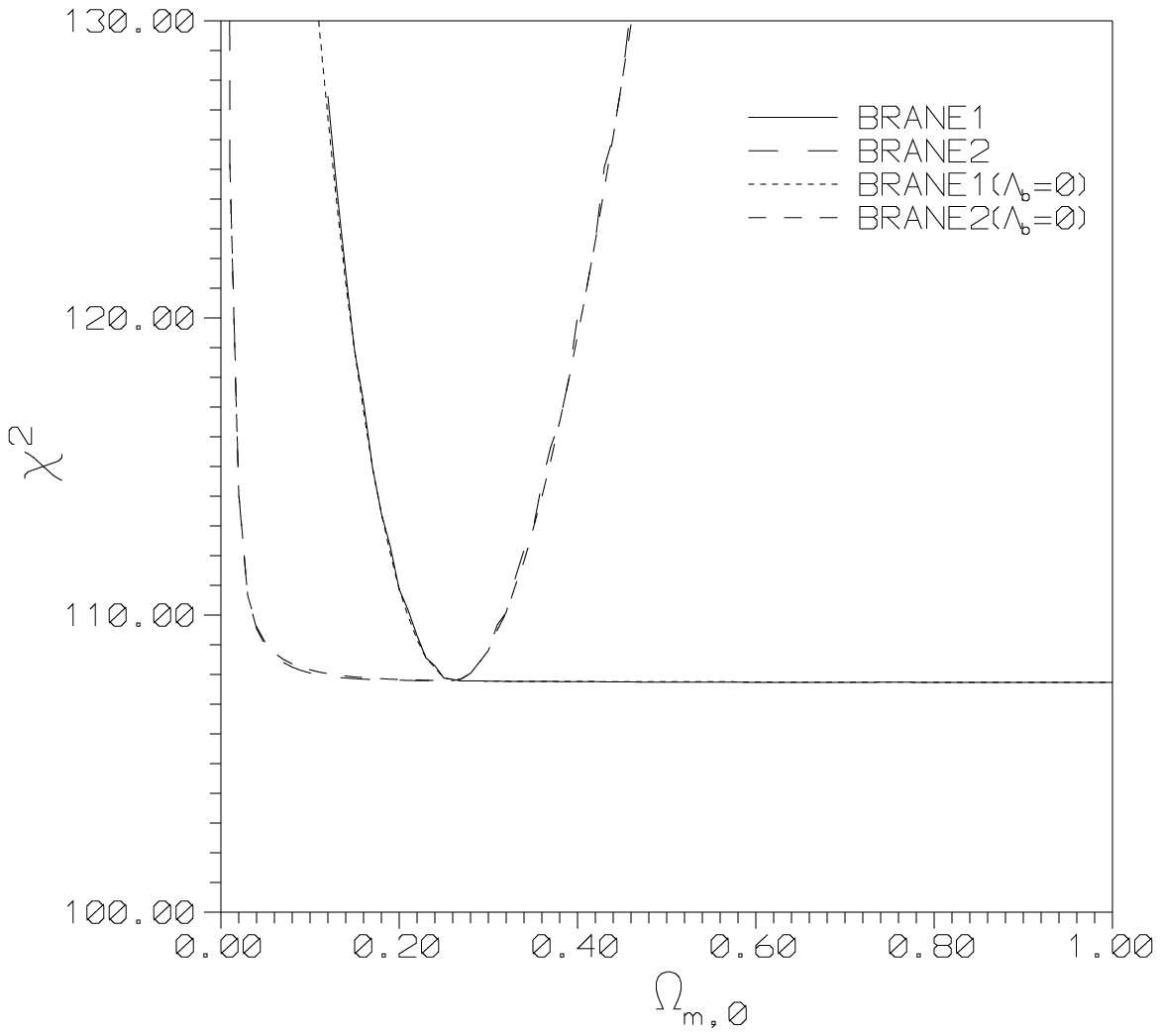}
\caption{The values of
$\chi^2$ with respect to the value of
$\Omega_{\text{m},0}$ for considered models, marginalized over remaining model
parameters. The left panel was obtained with the Gold sample while right panel
was obtained with Astier et al.'s sample.
We find no difference between $\chi^2$ values of the SSh ``Brane (1)'' model
and the $\Lambda$CDM model when $\Omega_{\text{m},0} \le 0.31$
($\Omega_{\text{m},0} \le 0.26$ for the Astier et al. sample). The opposite
case is for the SSh ``Brane (2)'' model where it does not differ from the
$\Lambda$CDM when $\Omega_{\text{m},0} >0.3$ ($\Omega_{\text{m},0} > 0.26$
for Astier et al.sample).
Please also not that for the Astier et al. sample SSh ``Brane'' models
are inseperable form SSh($\Lambda_b=0$) models.}
\label{fig:3}
\end{figure}

\begin{figure}
\includegraphics[width=0.45\textwidth]{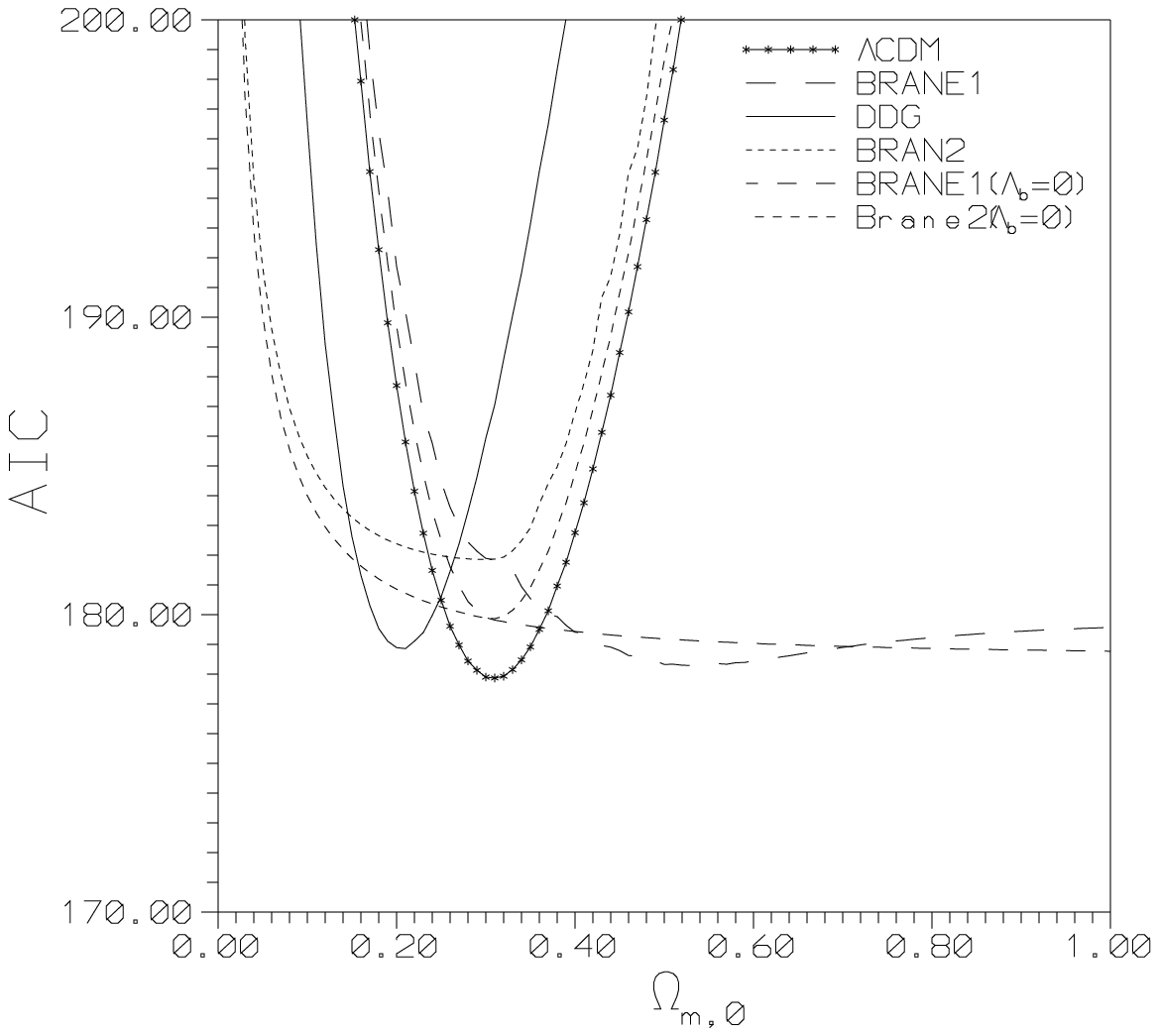}
\includegraphics[width=0.45\textwidth]{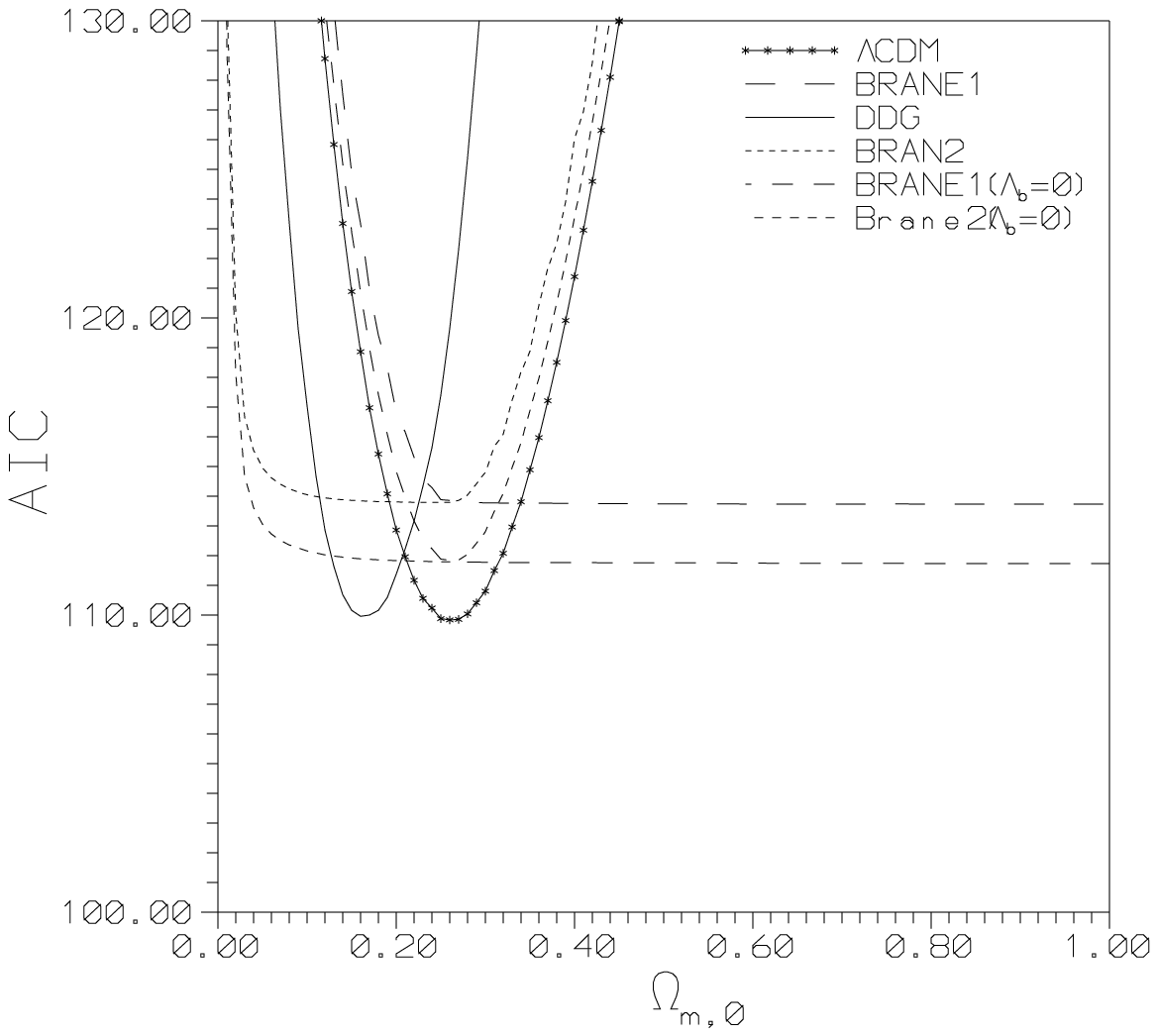}
\caption{The values of the AIC for the Gold sample from Riess et al.
(left panel) and Astier et al.'s sample (right panel) with respect to
the value of $\Omega_{\text{m},0}$ for the considered models.}
\label{fig:4}
\end{figure}

\begin{figure}
\includegraphics[width=0.45\textwidth]{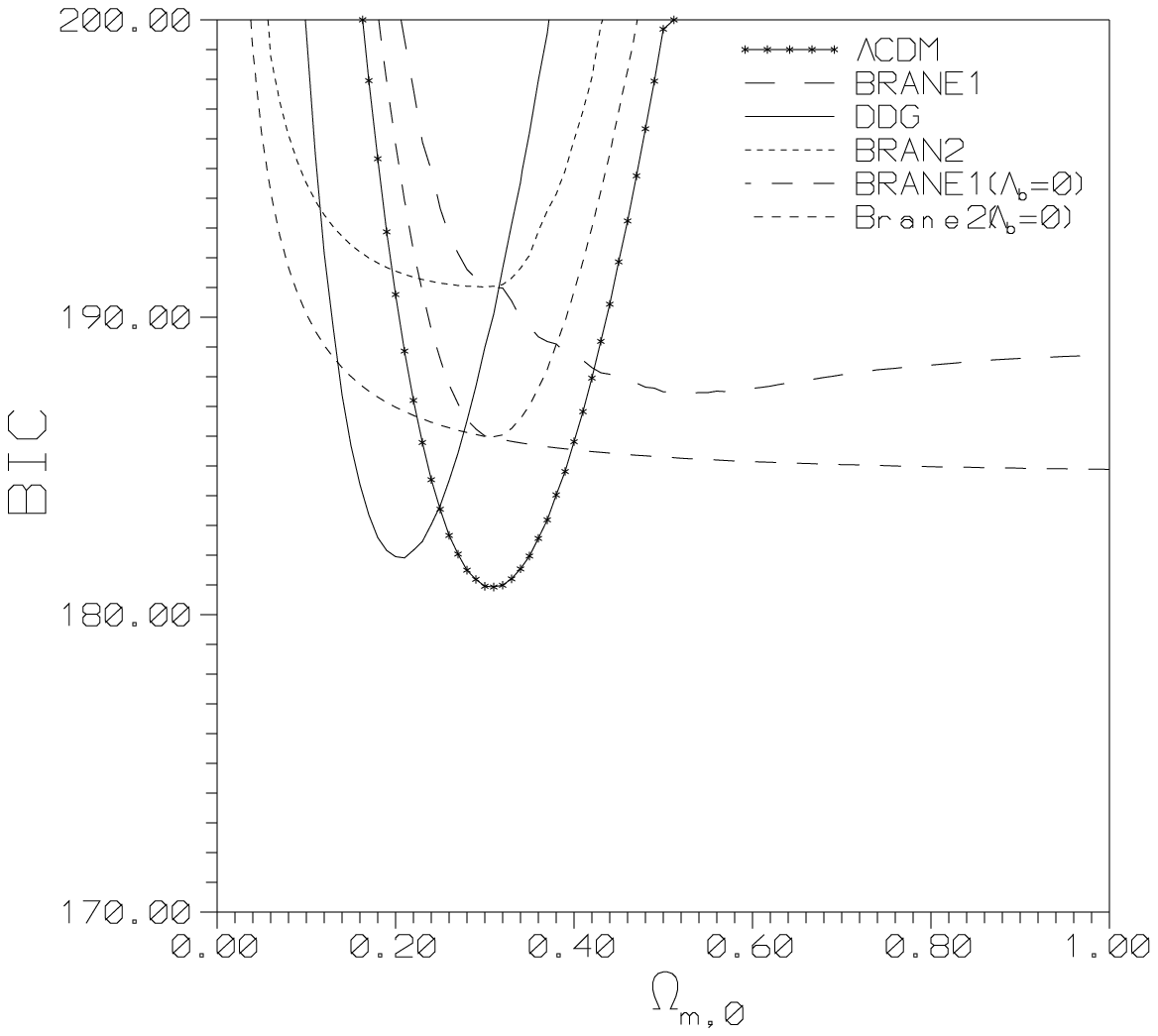}
\includegraphics[width=0.45\textwidth]{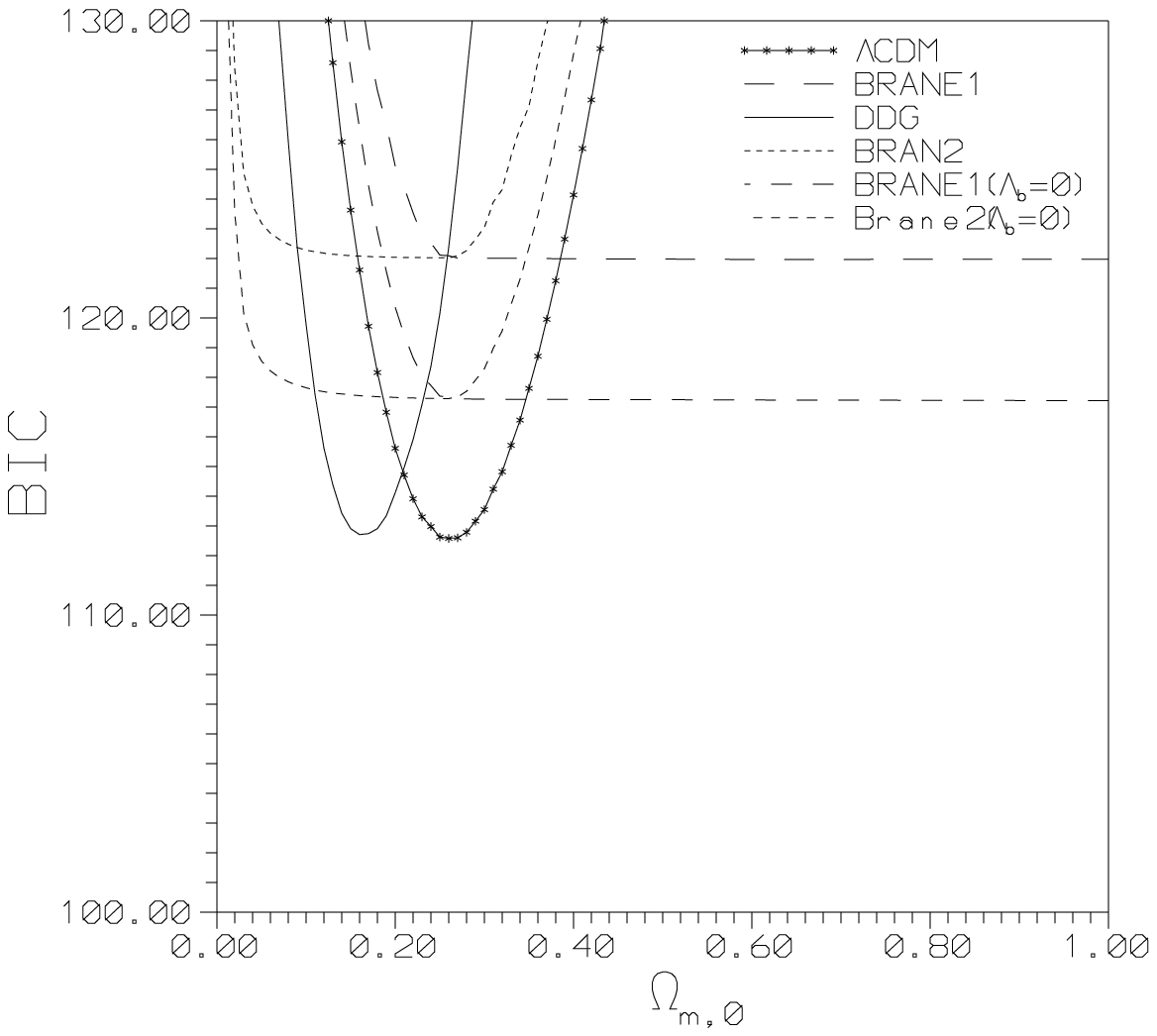}
\caption{The values of the BIC for the Gold sample from Riess et al.
(left panel) and Astier at al.'s sample (right panel) with respect to the value of $\Omega_{\text{m},0}$ for
the considered models.}
\label{fig:5}
\end{figure}

For deeper statistical analysis dependence of $\chi^2$
value, the AIC and BIC on $\Omega_{\text{m},0}$ is presented in
Fig.~\ref{fig:3}--\ref{fig:5}. These figures was obtained fixing
$\Omega_{\text{m},0}$ and than calculating $\chi^2$ value, the AIC and BIC
quantities for all points separately. Please note that if we fix
$\Omega_{\text{m},0}$, then the numbers of the free models parameters is
less by one.
 
In Fig.~\ref{fig:3} we present $\chi^2$ values with respect to fixed
$\Omega_{\text{m},0}$ for the Riess et al. and Astier et al. samples.
We find no difference between $\chi^2$ values of the SSh ``Brane (1)'' model,
SSH1($\Lambda_b=0$) model and the $\Lambda$CDM model when
$\Omega_{\text{m},0} \le 0.31$ ($\Omega_{\text{m},0} \le 0.26$ for Astier et al.
sample) while this difference diverges for greater $\Omega_{\text{m},0}$.
The opposite situation is for the SSh ``Brane (2)'' model where it differs
from the $\Lambda$CDM and SSH2($\Lambda_b=0$) in $\chi^2$ for
$\Omega_{\text{m},0} \le 0.3$ and it has no differences when
$\Omega_{\text{m},0} > 0.3$ ($\Omega_{\text{m},0} > 0.26$ for the
Astier et al. sample).
 
From this analysis we of course obtain the conclusion made by Alam and Sahni
\cite{Alam02} that the SSh model fits the SNIa data better than
the $\Lambda$CDM model (with exception $\Omega_{\text{m},0} \simeq 0.3$ for
the Gold sample and $\Omega_{\text{m},0} \simeq 0.26$  for the Astier et al. sample.
Please note that only for the SSh ``Brane (1)'' model analysed with the Gold
sample we obtain that the value of $\chi^2$ is significantly lower than for
the $\Lambda$CDM model.
 
Results obtained with both Riess et al. and Astier et al. samples are similar.
The main difference lies in best fitted values for $\Omega_{\text{m},0}$.
Also results obtained from the Riess et al. sample show the advantage of the SSh
``Brane (1)'' model over SSh1($\Lambda_b=0$) model for $\Omega_{\text{m},0}>0.31$
while from the Astier et al. sample do not show such preferences.
 
When we analyse ``Brane (1)'' model with the Riess et al. SNIa sample,
in Fig.~\ref{fig:3}--\ref{fig:5} one can observe the characteristic ``knee''
for the value $\Omega_{\text{m},0}\simeq 0.3$.
This effect comes from the fact that statistical analysis
of the SSh ``Brane (1)'' model for $\Omega_{\text{m},0}<0.31$
gives close to zero value of $\Omega_{\text{l},0}$ as a best fit.
It means that in this interval of $\Omega_{\text{m},0}$ the influence of
additional parameters is small or negligible. The analogous effect
takes place for the ``Brane (2)'' solution in the interval
$\Omega_{\text{m},0}>0.3$. Please note that only for the SSh ``Brane (1)''
model analysed with the Gold sample we obtain that the value of
$\chi^2$ is significantly lower than for the $\Lambda$CDM model.
Please note that Alam and Sahni (see Fig.~1 in \cite{Alam05})
did not find ``knee'' behaviour for ``Brane (1)'' model. The reason is
that in our analysis the parameter $\Omega_{\Lambda_b}$ is free while
in Alam and Sahni's paper this parameter is estimated and then fixed.
 
The most recent WMAP data \cite{Spergel06} seem to prefer the matter density
about $\Omega_{\text{m},0} \simeq 0.24$ which is lower than canonical
$\Omega_{\text{m},0}=0.3$. In this case we obtain that the
lowest value of $chi^2$ we obtain for ``Brane (2)'' model. Moreover
when we analyse the Riess et al. sample the value of $\chi^2$ for both
$\Lambda$CDM and DDG models are equal.
 
This preference of the SSh model over the $\Lambda$CDM model is not confirmed
by information criteria. With both the AIC and BIC criteria we obtain that
the model which minimizes both the AIC and BIC is the $\Lambda$CDM model.
There is also a significant difference between predictions of these models.
The $\Lambda$CDM model prefers a universe with $\Omega_{\text{m},0}$ close
to $0.3$, the SSh ``Brane (1)'' model favours a high density universe,
while the DDG model and the SSh ``Brane (2)'' model favour a low density
universe. In Fig.~\ref{fig:4} and Fig.~\ref{fig:5}  we present values of the
AIC and BIC for the considered models.
If $\Omega_{\text{m},0} \in (0.15,0.24)$ then the information criteria favour
the DDG model, while for $\Omega_{\text{m},0}<0.15$
($\Omega_{\text{m},0}<0.11$ from BIC) the SSh ``Brane (2)'' model is favoured.
For $\Omega_{\text{m},0} \in (0.24, 0.37)$, the $\Lambda$CDM is favoured while
for $\Omega_{\text{m},0}>0.37$ ($\Omega_{\text{m},0}>0.42$ when the
BIC is considered) the SSh ``Brane (1)'' model is preferred over $\Lambda$CDM
model. However, let us note that the value of $\Omega_{\text{m},0} \ge 0.4$
seems to be too high in comparison with the present extragalactic data
\cite{Peebles:2002gy}. When we analysed Astier et al. sample these values a
little change because of difference in best fitted values for
$\Omega_{\text{m},0}$ especially for $\Lambda$CDM model. For example
the $\Lambda$CDM model is favoured for $\Omega_{\text{m},0} \in (0.21,0.31)$
when the AIC is considered and for $\Omega_{\text{m},0} \in (0.21,0.35)$
when we consider BIC.

The BIC (and also the AIC in the case of Astier et al. sample) show preferences
the SSh1($\Lambda_b=0$) model over the SSh ``Brane (1)''. Both BIC and AIC
show preferences the SSh2($\Lambda_b=0$) model over the SSh ``Brane (2)''.
However the information
criteria still favour the $\Lambda$CDM over SSh1($\Lambda_b=0$) model if
$\Omega_{\text{m},0}<0.36$ in the case of the AIC ($\Omega_{\text{m},0}<0.31$
for  Astier et al. sample) and $\Omega_{\text{m},0}<0.4$
($\Omega_{\text{m},0}<0.35$ for Astier sample) in the case of the BIC.

With no prior on the value of matter content $\Omega_{\text{m},0}$
the information criteria favour the $\Lambda$CDM model rather than
the FRW brane models with extra dimensions. The same result in favour
of the $\Lambda$CDM model is obtained if $\Omega_{\text{m},0}$ is not
significantly different from the canonical value $\Omega_{\text{m},0}=0.3$.
However, taking the other value of $\Omega_{\text{m},0}$ from independent
observations (e.g. recent WMAP data --- $\Omega_{\text{m},0} \simeq 0.24$
\cite{Spergel06}) appears to equally favour the $\Lambda$CDM and DDG
models (also SSh2($\Lambda_b=0$) if we consider AIC only)  with the Gold
Riess SNIa sample. Please also note that if allowed
non flat models, non flat $\Lambda$CDM and DDG models are again equally
favoured by information criteria \cite{Szydlowski:2005}.
 
\begin{table}
\noindent \caption{The values of the $\chi^2$, the AIC and BIC for the models
from Table~\ref{tab:1}.}
\label{tab:2}
\begin{tabular}{c|ccccc|ccccc}
\hline \hline
\multicolumn{1}{c}{}&
\multicolumn{5}{c}{Riess et al.}&
\multicolumn{5}{c}{Astier et al.}\\
\hline
case & $\chi^2$ & AIC &  BIC  &$\Omega_{\text{m},0}$& $\chi^2$/dof& $\chi^2$ & AIC &  BIC &$\Omega_{\text{m},0}$& $\chi^2$/dof
 \\
\hline
1  & 175.9  & 179.9  & 186.0 & 0.31 & 1.135 & 107.8  & 111.8  & 117.3 & 0.26 & 0.954 \\
2  & 176.9  & 180.9  & 187.0 & 0.20 & 1.141 & 108.0  & 112.0  & 117.5 & 0.17 & 0.956 \\
3a & 172.3  & 180.3  & 192.5 & 0.53 & 1.126 & 107.7  & 115.7  & 126.7 & 0.51 & 0.970 \\
3b & 175.8  & 183.8  & 196.0 & 0.30 & 1.149 & 107.8  & 115.7  & 126.8 & 0.26 & 0.971 \\
4a & 174.8  & 180.8  & 189.9 & 1.00 & 1.135 & 107.7  & 113.7  & 121.9 & 1.00 & 0.962 \\
4b & 175.9  & 181.9  & 191.0 & 0310 & 1.142 & 107.8  & 113.8  & 122.0 & 0.26 & 0.962 \\
\hline
\end{tabular}
\end{table}
 
\begin{table}
\caption{The values of the
$\chi^2$, the AIC and BIC for the models from Table~\ref{tab:1} with the
prior $\Omega_{\text{m},0}=0.2,0.3,0.4$ obtained with the Gold from
Riess et al. (denoted as $G$) and Astier et al. (denoted as $A$) samples.}
\label{tab:3}
\begin{tabular}{c|cccc|cccc|cccc}
\hline \hline
\multicolumn{1}{c}{}&
\multicolumn{4}{c}{$\Omega_{\text{m},0}=0.2$}&
\multicolumn{4}{c}{$\Omega_{\text{m},0}=0.3$}&
\multicolumn{4}{c}{$\Omega_{\text{m},0}=0.4$}\\
\hline \hline
case & $\chi^2$ & AIC & BIC & $\chi^2$/dof & $\chi^2$ & AIC & BIC & $\chi^2$/dof& $\chi^2$ & AIC & BIC & $\chi^2$/dof\\
\hline
G1  & 185.7 & 187.7 & 190.8 & 1.190 & 175.9 & 177.9 & 181.0 & 1.128 & 180.8 & 182.8 & 185.9& 1.159\\
G2  & 176.9 & 178.9 & 182.0 & 1.134 & 183.9 & 185.9 & 189.0 & 1.179 & 200.0 & 202.0 & 205.1& 1.282\\
G3a & 185.7 & 191.7 & 200.8 & 1.205 & 175.9 & 181.9 & 191.0 & 1.142 & 173.4 & 179.4 & 188.5& 1.126\\
G3b & 176.5 & 182.5 & 191.6 & 1.146 & 175.8 & 181.8 & 190.9 & 1.142 & 180.8 & 186.8 & 195.9& 1.174\\
G4a & 185.7 & 189.7 & 195.8 & 1.198 & 175.9 & 179.9 & 186.0 & 1.135 & 175.4 & 179.4 & 185.5& 1.131\\
G4b & 176.9 & 190.9 & 197.0 & 1.141 & 175.9 & 179.9 & 186.0 & 1.135 & 180.8 & 184.8 & 190.9& 1.166\\
\hline
A1  & 110.9 & 112.9 & 115.6 & 0.973 & 108.8 & 110.8 & 113.5 & 0.954 & 119.4 & 121.4 & 124.1& 1.047\\
A2  & 109.4 & 111.4 & 114.1 & 0.960 & 130.7 & 132.9 & 135.4 & 1.147 & 181.1 & 183.1 & 185.8& 1.589\\
A3a & 110.9 & 116.9 & 125.1 & 0.990 & 107.8 & 113.8 & 122.0 & 0.962 & 107.8 & 113.8 & 122.0& 0.962\\
A3b & 107.8 & 113.8 & 122.0 & 0.963 & 108.8 & 114.8 & 123.0 & 0.971 & 119.4 & 125.4 & 133.7& 1.066\\
A4a & 110.9 & 114.9 & 120.4 & 0.981 & 107.8 & 111.8 & 117.2 & 0.954 & 107.8 & 111.8 & 117.3& 0.954\\
A4b & 107.8 & 111.8 & 117.3 & 0.954 & 108.8 & 112.8 & 118.3 & 0.963 & 119.4 & 123.4 & 118.9& 1.057\\
\hline
\end{tabular}
\end{table}

Recently Eisenstein et al. have analysed the baryon oscillation peaks (BOP)
detected in the SDSS Luminosity Red Galaxies Survey \cite{Eisenstein:2005}.
They found that value of $A$
 
\begin{equation}
\label{eq:16}
A \equiv =\frac{\sqrt{\Omega_{\text{m},0}}}{E(z_1)^{\frac{1}{3}}}
\left(\frac{1}{z_1\sqrt{|\Omega_{k,0}|}}
\mathcal{F} \left( \sqrt{|\Omega_{k,0}|} \int_0^{z_1} \frac{d z}{E(z)} \right)
\right)^{\frac{2}{3}}
\end{equation}
where $E(z) \equiv H(z)/H_0$ and $z_1=0.35$ is equal $A=0.469 \pm 0.017$.
The quoted uncertainty corresponds to one standard deviation, where a
Gaussian probability distribution has been assumed. This constraints could
also be used for fitting cosmological parameters
\cite{Astier:2005,Fairbairn:2005,Alam05,Maartens06,Guo06}. The  cosmological parameters are
derived from Bayes' theorem in the standard way \cite{Riess:1998cb}.
Estimated value of $\Omega_{\text{m},0}$ as well as confidence levels
intervals for $\Omega_{\text{m},0}$ we obtain by marginalizing the
probability density functions over remaining parameters, assuming uniform priors.
The estimated value of the model parameter $\Omega_{\text{m},0}$ as well
as 95\% confidence interval of $\Omega_{\text{m},0}$ from the BOP is
presented in Table~\ref{tab:4}. Please note that for both SSh ``Brane (1)''
and SSh ``Brane (2)'' models, the 95\% confidence level regions (obtained with
the BOP analysis) and the regions of $\Omega_{\text{m},0}$ at which these models
are favoured by information criteria, are  disjoint. Therefore from the joint
analysis of the BOP and information criteria we obtain that the $\Lambda$CDM
model is still favoured over both the SSh ``Brane (1)'' and ``Brane (2)''
models. Our analysis also confirm previous conclusion in
Ref.~\cite{Fairbairn:2005} that the flat DDG model can be virtually ruled out
by statistical analysis. Our analysis allowed only that the SSh1($\Lambda_b=0$)
model could be preferred over $\Lambda$CDM model if
$\Omega_{\text{m},0} \simeq 0.4$.
 
\begin{table}
\noindent \caption{The constraints for
$\Omega_{\text{m},0}$ from the baryon oscillation peak test for the models
from Table~\ref{tab:1}. We present best fitted value as well as 95\%
confidence interval.}
\label{tab:4}
\begin{tabular}{c|cc}
\hline \hline
case & $\Omega_{\text{m},0}$& 95\% level \\
\hline
1  &  0.273 &  (0.228,0.326)\\
2  &  0.305 &  (0.252,0.363)\\
3a &  0.23  &  (0.16,0.28)\\
3b &  0.32  &  (0.25,0.49)\\
4a &  0.298 &  (0.251,0.358)\\
4b &  0.367 &  (0.297,0.498)\\
\hline
\end{tabular}
\end{table}
 
\section{Conclusion}
 
The main goal of this letter is to decide which class of models with dark
energy are distinguished by statistical analysis of SNIa data. For this aim
the Akaike and Bayesian information criteria are adopted. Two categories of
the models were considered, one with dark energy in the form of fluid
violating strong energy condition and second in which dark energy is the
present manifestation of embedding a brane (our universe) in a larger, higher
dimensional bulk space. One concludes that both the AIC and BIC
weigh in favour of the models of the first category (with the $\Lambda$CDM as
their representative case) over the FRW brane model with extra dimensions.
Assuming the prior $\Omega_{\text{m},0}=0.3$ both the AIC and BIC
weigh in favour of the $\Lambda$CDM model.
However please note the most recent WMAP data \cite{Spergel06} seem to
favour a lower value of matter density $\Omega_{\text{m},0} \simeq 0.24$.
For such value of $\Omega_{\text{m},0}$ at least during the
analysis with the Gold Riess at al. SNIa sample both the AIC and BIC equally
favour the $\Lambda$CDM and DDG models. Please also note that if allowed
non flat models, the non flat $\Lambda$CDM and DDG models are again equally
favoured by the information criteria \cite{Szydlowski:2005}.
 
The further conclusions are the following.
\begin{itemize}
\item If we consider models in which all model parameters are fitted then the
$\Lambda$CDM model is preferred.  DDG model give a  greater value of $\chi^2$
over the $\Lambda$CDM  model. Both the SSh ``Brane (1)'',``Brane (2)'' as well
as SSh($\Lambda_b=0$) models, under the similar quality of the fit as for the
$\Lambda$CDM model, contains more parameters than the $\Lambda$CDM model.
\item When we consider the prior on $\Omega_{\text{m},0}$ then for
$\Omega_{\text{m},0} < 0.24$ the DDG model is favoured by the information criteria
over the $\Lambda$CDM model while for the ``normal'' density universe with
$\Omega_{\text{m},0} \simeq 0.3$ the $\Lambda$CDM model is favoured.
Only for the high density universe ($\Omega_{\text{m},0}>0.38$ from the AIC
and $\Omega_{\text{m},0}>0.42$ from the BIC) is the SSh ``Brane (1)'' model
preferred over the $\Lambda$CDM model. The SSh ``Brane (2)'' model is
preferred for the low density universe ($\Omega_{\text{m},0}<0.15$ from the AIC
and $\Omega_{\text{m},0}<0.11$ from the BIC).
When we analyse Astier et al. sample these values a little change.
The $\Lambda$CDM model is favoured for $\Omega_{\text{m},0} \in (0.21,0.31)$
when we consider AIC  and for $\Omega_{\text{m},0} \in (0.21,0.35)$ when we
consider BIC.
\item The BIC information criterion favours the SSh1($\Lambda_b=0$)  model  over
the $\Lambda$CDM for $\Omega_{\text{m},0}>0.4$ ($\Omega_{\text{m},0}>0.36$ for
the Astier et al. sample). The AIC favour the SSh1($\Lambda_b=0$)  model even for
$\Omega_{\text{m},0}>0.36$ ($\Omega_{\text{m},0}>0.31$ for the Astier et al.
sample).
\item If we compare SSh ``Brane '' models with SSh($\Lambda_b=0$) models using
the BIC than we obtain that the SSh($\Lambda_b=0$) model is preferred over SSh
``Brane '' models.
\item If we consider the 95\% confidence level regions obtained with the BOP
analysis and the region of $\Omega_{\text{m},0}$ where  model ``Brane (1)''
and ``Brane (2)'' are favoured by the information criteria we find that they
are disjoint. It means that the $\Lambda$CDM model is still favoured over
both SSh ``Brane (1)'' and ``Brane (2)'' models from joint analysis.
\item We find that the $\Lambda$GDP model is preferred over the $\Lambda$CDM
model by the joint analysis but only if $\Omega_{\text{m},0}$ is close to
$0.4$.
\end{itemize}
 
We find no difference between $\chi^2$ values of the SSh ``Brane (1)'' model
and the $\Lambda$CDM model when $\Omega_{\text{m},0} \le 0.31$
($\Omega_{\text{m},0} \le 0.26$ for the Astier et al. sample). The opposite
situation is for the SSh ``Brane (2)'' model where and it has no differences
when $\Omega_{\text{m},0} >0.3$ ($\Omega_{\text{m},0} > 0.26$ for the Astier et al.
sample). In these intervals the SSh ``Brane (1)'' and SSh ``Brane (2)'' models
are, therefore, indistinguishable from the concordance $\Lambda$CDM model. And
we have some kind of dynamical equivalence between two pairs of models in
the above found intervals. This lack of differentiation in terms of $\chi^2$
can be overcome by the Akaike and Bayesian information criteria
(see Fig.~\ref{fig:4} and Fig.~\ref{fig:5}).
 
In observational cosmology we are encountered with the so-called degeneracy
problem which consists in the existence of many theoretical models with
dramatically different cosmological scenarios (big bang versus bounce,
big rip versus de Sitter phase, etc.) but in a good agreement with the current
observations. A nice way of overcoming this problem seems to be
adopting the information criterion approach. Since it provides a very
simple and objective criterion for the inclusion of additional
parameters in the cosmological model, it could be used for model
selection instead of the best-fit method.
In context of dark energy the information criteria give us information
if the present observational data suggest taking into account new degrees
of freedom. Of course in any case introducing a new term can be suggested
in the other way -- for example from theoretical prediction.
 
Our general conclusion is that the high precision detection of distant
type Ia supernovae could justify an inclusion of new parameters related with
embedding our Universe in bulk space. Our results were obtained from SNIA
data set and baryon oscillation peak. However, to make the final decision
which model describes our Universe it is necessary to obtain the precise
value of $\Omega_{\text{m},0}$ from independent observations. Other future
investigations such as gravitational lensing, WMAP, X-ray gas mass fraction
measurements are required for the final resolution of the problem.
 
\ack{M. Szydlowski acknowledges the support by KBN grant no. 1 P03D 003 26.
Authors thanks Dr Riess and Dr Astier for explanation in details his supernovae
samples. The authors are very grateful to Yuri Shtanov and Varun Sahni for
helpful comments and remarks. We also thanks the anonymous referee for comments
which help us to improve significantly the original version of the paper.}

\end{document}